# A Hybrid Approach to Domain-Specific Entity Linking

Alex Olieman          Jaap Kamps          Maarten Marx          Arjan Nusselder

University of Amsterdam, Amsterdam, The Netherlands

{olieman|kamps|maartenmarx}@uva.nl          arjan@nusselder.eu

## ABSTRACT

The current state-of-the-art Entity Linking (EL) systems are geared towards corpora that are as heterogeneous as the Web, and therefore perform sub-optimally on domain-specific corpora. A key open problem is how to construct effective EL systems for specific domains, as knowledge of the local context should in principle increase, rather than decrease, effectiveness. In this paper we propose the hybrid use of simple specialist linkers in combination with an existing generalist system to address this problem. Our main findings are the following. First, we construct a new reusable benchmark for EL on a corpus of domain-specific conversations. Second, we test the performance of a range of approaches under the same conditions, and show that specialist linkers obtain high precision in isolation, and high recall when combined with generalist linkers. Hence, we can effectively exploit local context and get the best of both worlds.

## Categories and Subject Descriptors

H.3.1 [**Information Systems**]: Content Analysis and Indexing – *abstracting methods, indexing methods, linguistic processing*.

## Keywords

Information Extraction, Entity Linking, Semantic Annotation, Conversational Text, Minutes, Parliamentary Proceedings.

## 1. INTRODUCTION

In the Entity Linking (EL) task, textual mentions are linked to corresponding Knowledge Base (KB) entries. The majority of state-of-the-art EL systems utilize one or more open-domain KBs, such as Wikipedia, DBpedia, Freebase, or YAGO, as basis for learning their entity recognition and disambiguation models [10]. This approach shows definite merit when the target corpus consists of texts with heterogeneous topical contents [9], e.g. in a random sample of news articles or blog posts.

Anyone with the desire to annotate a domain-specific (i.e. homogeneous) corpus, however, will at some point face sub-optimal results when using a domain-agnostic EL system. This problem has been identified as one of three promising research directions in this area [10]. The main aim of this paper is to investigate domain-specific Entity Linking.

The straightforward solution of training the state-of-the-art recognition and disambiguation models on the corpus (instead of on Wikipedia) can be extremely costly if accurate training data needs to be handcrafted from scratch. Alternatively, a generalist EL system can be used on the corpus without modification. This is clearly a minimum-cost option, but its performance depends highly on the similarity of the corpus with the text that the models are based on (e.g. Wikipedia articles). Currently, the most practical approach for domain-specific EL likely lies somewhere in the middle: some adaptation needs to occur, and preferably with minimum effort. In this paper, we propose to use specialized linkers for salient entity types within the corpus' domain, which can work in concert with a generally trained model.

We apply our approach to conversational text, in particular parliamentary proceedings, i.e. the minutes of parliamentary debates. When generating semantic annotations from conversational records, e.g. minutes or online conversations, the structure of the records already provides much useful information. It tells us, for instance, who was the speaker of each unit of speech, who spoke in response to whom, and who participated in the conversation. Additional information may be provided by metadata for each conversation, such as when and where it took place, between which group of people, or what the occasion or agenda was. Moreover, when the structure of the records is congruent, parsing this information is straightforward.

This offers a springboard for generating valuable annotations by applying subsequent NLP to the full records. This paper focuses on utilizing Information Extraction (IE) techniques–EL in particular–to enrich existing structure-based annotations. The techniques under investigation in this paper are designed to be applicable to written records of any kind of conversation.

Our contribution lies in answering the following questions:

1. How can mentions of the most salient entity types within a corpus be linked at a low cost in terms of system development and domain expertise?
2. How to construct a reusable benchmark for EL on conversations, that allows comparison between systems and combinations of systems?
3. How effective are the specialist linkers, and how effective is their hybrid combination with generalist EL systems?

## 2. RELATED WORK

Until the beginning of the 21st century, it was common to collect the domain knowledge that was needed for an IE task in a KB [9]. Progress in supervised machine learning, and the availability of high-coverage encyclopedic resources, however, has led to the use of open-domain KBs in recent years. The domain-specific nature of IE is no longer expressed in the KB, but instead in the training data [9]. This has moved the adaptation cost of applying EL on a specific corpus from the system developer to the domain expert.

Efforts to reduce the need for domain experts have been made by semi-supervised adaptation of generalist models to a target corpus [10]. One promising direction is Transfer Learning, which is known to work for classification tasks [5], whereas this has not been demonstrated sufficiently for EL. Alternatively, a domain-relevant part of the KB can be selected by excluding KB-entries that are more likely to be generated by a parsimonious unigram model of the KB (with the corpus as background), than by the unigram corpus model [2]. In Berlanga et al. [2], KB-entries are also tailored by basing entity-specific language models on both the corpus and the KB.

The recently presented GERBIL [11] is a KB-agnostic EL benchmarking framework, which addresses issues with the comparability and reproducibility of EL systems and experiments. Our benchmark is complementary to GERBIL, in that it additionally allows combinations of EL systems to be evaluated.

## 3. DOMAIN-SPECIFIC LINKING

Our approach is to develop specialist linkers for entity types that are mentioned frequently in the target corpus. These linkers capitalize on a small amount of background knowledge, and achieve entity recognition and disambiguation by means of pattern detection, string matching, and structured queries against the corpus.

We have selected the Dutch parliamentary proceedings as the target corpus for an experiment, available in an XML format with rich (structural) annotations, and which covers 1814 until today. The automated analysis of parliamentary proceedings is part of a larger international effort, and has been facilitated by previous work in the PoliticalMashup project [7].

Two off-the-shelf EL systems are used as baseline systems, and also as components for our combination approach. The first is DBpedia Spotlight v0.7, which takes raw text as input and produces links with generative models based on DBpedia and Wikipedia [4]. It distinguishes itself from other state-of-the-art EL systems by creating entity-specific language models from the context of Wikipedia page links, rather than from the pages themselves. The second system is comprised of separate entity recognition and disambiguation modules. Frog is an NLP workbench for Dutch [3], from which the phrase chunking module is used to identify noun phrases. The identified phrases are subsequently passed on to the UvA Semanticizer, which takes a learning to re-rank approach to disambiguation [8].

### 3.1 Domain-specific candidate entities

The simplest way that we have considered to annotate entities of a specific type starts by collecting names for the entities in question, including acronyms. These names are stored in a dictionary, which maps them to canonical URIs. Subsequently, a state machine that encodes all names is constructed by the Aho-Corasick algorithm [1]. This allows the set of names to be matched in an arbitrary input string, and the URI of mentioned entities to be found in the dictionary.

This minimal-effort approach is fundamentally limited to entity types in which no ambiguity exists. The many-to-one mapping from names to URIs deals with synonymy, but does not allow a single name to be associated with multiple entities. Such a linker therefore needs to target a type with few instances, or in which ambiguous names are already avoided because they would confuse communication. It is difficult, for instance, to find brands in the same sector that share a name or acronym.

In our corpus we target Dutch political parties ($n=155$), because they are highly relevant as well as unambiguously named. The linker uses case-insensitive, leftmost-longest string matching. Any matches that are part of a longer token are rejected. The second prerequisite for such a simple approach is that none of the entity names should also occur as common words. If such names do occur, this may be addressed with case-sensitive string matching.

### 3.2 Genre-specific characteristics

The genre of conversational text exhibits several characteristics that can act as useful clues for EL. We focus on features that relate to mentioned persons; a salient entity type in many kinds of conversation. Conversations have a *temporal aspect*, i.e. all words have been spoken or written at some point in time, and it is likely for a longitudinal corpus that the frequency with which a particular person is mentioned varies over time. Conversations are also *situated*: they occur in a (virtual) space in which a person may be present or not. These features can be used for disambiguation.

If a corpus focuses on a specific domain of discourse, there may also be characteristic ways in which names are used, as in etiquette and/or jargon. Government and parliament members (in short: *members*) adhere to guidelines on how to address each other during a parliamentary debate. Members are addressed as, e.g., Mr., Mrs., colleague, minister, or secretary of state. This characteristic of the corpus can be utilized by detecting *where* a member is mentioned, and thereby avoid the ambiguity between their name and its homonyms. Government members are often mentioned only by their role (e.g. minister), which may be followed by a portfolio (e.g. the minister of finance). We have developed a regular expression that matches such patterns, and which avoids including words that are not part of a name or portfolio.

The phrases that are found by the regular expression need to be linked to the URIs of the mentioned members ($n=3,664$). In any parliamentary debate, most of the people that are mentioned are present in that session. The PoliticalMashup proceedings include a structured speakers list which we use to resolve such mentions. We use an index of members to disambiguate mentions of non-speakers, and query it with their name, and the date and the house in which the debate took place. A link is generated only if this query has a single result, i.e., when it can be created with high confidence.

We have also built an index of government members by time period, role, and portfolio, with which to resolve mentions by role. If the portfolio is not mentioned, the linker assumes that the mentioned person is a speaker. The speakers list is searched for any members with the mentioned role (i.e. minister or secretary). If there are multiple candidates, we assume that the last-mentioned member with this role is mentioned here.

## 4. EVALUATION

In this section we describe the development of a reusable benchmark for EL on a corpus of domain-specific conversations. Our approach–using specialist linkers for salient entity types, and their combination with general-purpose EL systems–is tested with this benchmark, and we report on its results.

### 4.1 Benchmark

We have selected a sample of Dutch parliamentary proceedings from the period 1999-2012. In consideration of the uneven spread in topical content over the various debates, we have stratified the sample into governmental departments, with which we assume the topical content is strongly associated. There is no formal one-to-one relation between debates and departments, and therefore we have used speakers with a government position as an indicator.

The size of the sample is restricted to the approximate length of a 3-hour debate, to limit the amount of time that our volunteer annotators needed to spend on manual annotation. From this overall limit, we allocated per-department quota in proportion to the number of debates associated with the department during the full period. For each department, a random debate is selected and taken out of the pool. From this debate, a random *scene*–a single member's speaking time with optional interruptions and replies–is picked, and included in the sample. These steps are repeated for all departments in round-robin fashion until the overall limit is reached. Departments for which the quotum is full skip their turn.

Table 1. Composition of the stratified sample

| Department | \|scenes\| | \|a\|[1] | \|a_per\| | \|a_org\| |
|---|---|---|---|---|
| Economic Affairs | 4 | 97 | 29 | 10 |
| Security and Justice | 4 | 90 | 31 | 7 |
| Infrastructure and the Environ. | 4 | 79 | 41 | 14 |
| Without department | 4 | 72 | 33 | 16 |
| Social Affairs and Employment | 4 | 61 | 32 | 10 |
| Interior and Kingdom Relations | 4 | 57 | 17 | 11 |
| Finance | 4 | 53 | 30 | 1 |
| Foreign Affairs | 3 | 51 | 7 | 5 |
| Education, Culture and Science | 2 | 43 | 16 | 7 |
| Health, Welfare and Sport | 4 | 32 | 19 | 1 |
| General Affairs | 3 | 32 | 11 | 5 |
| Defense | 3 | 15 | 5 | 4 |
| *Total* | 43 | 682 | 271 | 91 |

This sample, see Table 1, was subsequently annotated by the two baselines and the specialist linkers. The resulting annotations were pooled into the sample's XML format. In order to assess the quality of these annotations against a consistent *gold standard*, we employ two human annotators for an independent and a consensus-building annotation round. We have established guidelines for them, e.g.: adjectively used names should be linked, but metaphorical speech and pronouns should not.

We have additionally developed a web interface to facilitate the creation of the gold standard by human annotators. The interface displays a single debate at a time, and clearly marks the scene of interest. The phrases that have been annotated by at least one of the systems are highlighted in this scene, and the annotator is able to select the mentioned entity from a list, or by entering a Wikipedia or PoliticalMashup URL manually. The annotator may also indicate that the mentioned entity is not present in either KB, or that the phrase should not be annotated at all. This benchmark does not evaluate entity mention boundaries in the interest of simplifying the manual annotation task. Overlapping annotations are displayed by the interface as their longest span, and annotators are able to enter multiple valid URLs. The pre-selection of candidate entities is achieved by deduplicating the system annotations, and adding to this the top results from queries to Wikipedia and PM with the annotated phrase.

### 4.2 Combination of system annotations

We have taken a simple approach to combining the output of multiple systems to address the aim of linking mentioned entities that are specific to the domain, as well as other entities. This approach is intended not to make use of any training data.

Earlier work on how to combine the output of multiple generalist EL systems has used a voting method [6], and shows it to be somewhat effective. Taking a vote on how to link, however, seems less promising when systems are specialized towards certain entity types. If we take the analogy of asking a question in a room full of specialists, who answers the question matters a great deal. We therefore employ a preference ordering instead: the most specialized (i.e. estimated high-precision) system is asked to link a phrase first, and only if it doesn't the second system in the order is asked, and so on. By adding a generalist EL system at the end of the chain, the phrases that mention non-domain-specific entities also have their chance at being linked.

---

[1] Number of phrases that have been annotated by at least one system.

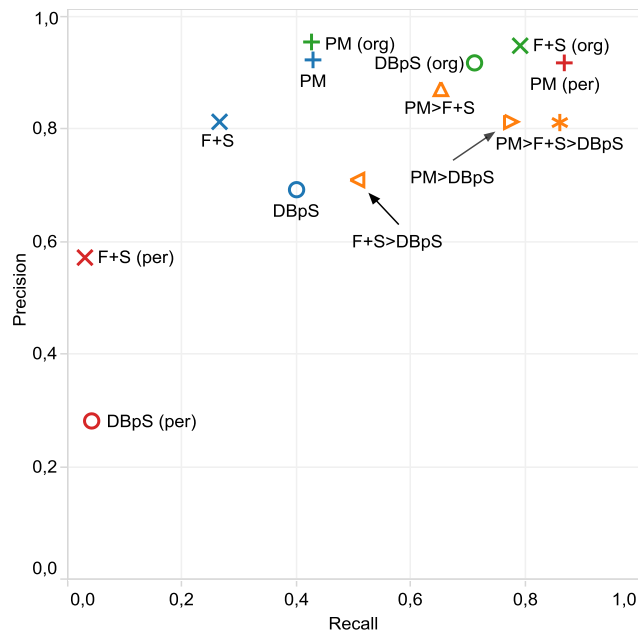

Figure 1. Performance of the EL systems and combinations

### 4.3 Results

We have used the developed benchmark to assess the correctness of the annotations that were generated by the specialist linkers and the baseline systems. To this end, we calculate precision and recall between the system and gold annotations ($n=639$). Figure 1 shows the performance of the specialist linkers (+, *PM*), DBpedia Spotlight (○, *DBpS*), Frog+Semanticizer (×, *F+S*), and preference-ordered combinations thereof (◁,△,▷,∗). For the single systems, the performance on annotations that link to persons and organizations is also shown separately.

These results show that the specialist linkers were able to generate a larger number of accurate annotations for the corpus than either of the baseline systems, whilst limited to two specific entity types. F+S is the more precise of the baselines, but DBpS produces a greater number of potentially useful links. Both baselines are not much good at identifying the people that are mentioned in this corpus, as we had expected, but F+S is surprisingly good at annotating organizations.

Specialist linkers, generally speaking, gain a head start over generalist systems by working with a smaller set of candidate entities. They are able to spot phrases that they should link with higher confidence, and in some cases lack the need to disambiguate, because they only know about a one-to-one mapping from the spotted phrase to an entity. Where generalist EL systems are somewhat biased towards entities with a high Web-presence, a specialist system should be biased towards entity types that are of interest to the users of a particular corpus. The linker with which we targeted parliament members is additionally empowered by some temporal awareness, and a mapping from government positions to office-holders. It is therefore the only system that can accurately link persons that are only mentioned by their office.

Our approach of combining a relatively simple custom-made EL system with an off-the-shelf EL system has also proven to be successful. Letting the specialist PM linkers annotate any phrases they could, and to let the remaining phrases be annotated by either DBpS or F+S, produced a significantly better result (+27% ≤ $\Delta F_1$ ≤ +99%) than any of the systems could by themselves. If high recall is of importance, it can be achieved by combining all three systems in an order of descending precision. The number of phrases for which only one of the combined systems produces an annotation gives an upper bound for the gain in recall. There are 548 of such phrases for DBpS and PM, 451 for F+S and PM, and 333 for DBpS and F+S in our corpus.

## 5. APPLICATIONS

The potential for semantic annotations to improve information access is clear when we focus on users with a deep interest in the corpus' domain. An obvious application is in semantic search [2], where entity linking can help address issues with homonymy and synonymy in document retrieval. More notably, entity links can simplify the kind of queries that are used in corpus analysis, to which the desired answer is not a list of documents.

Consider this example for the genre of conversational text: *give an overview of all the questions that have been addressed to person X*. This information need could be answered at a high level, e.g., by displaying a timeline which shows the frequency of asked questions, and, for any selected time period, who where the top question-askers and which other entities are mentioned frequently in the context of these questions. A user may also drill-down into a (filtered) concordance view of the questions addressed to person X. The advantage over keyword search is that EL can resolve partial and ambiguous name matches, and mentions of role-holders, to specific individuals. The way in which an entity is mentioned thus becomes part of the answer, instead of the query.

Another example is the application of EL for Social Network Analysis. When the conversational corpus is viewed as a social network, the structure of the conversations can already shed light on some of the relations in this network. In the parliamentary domain, e.g., it is possible to derive a graph of *who is interrupted by whom* from the structure of the proceedings [7]. Entity links allow us to see the much broader graph of *who mentioned whom* during a conversation. By showing this mention graph against the background of the interruption graph, it becomes easy to explore the cases in which people mention each other for other reasons than a direct reply.

Finally, the low-cost annotation approach that we have described can be used to bootstrap other EL approaches, and other Information Extraction tasks. In cases where it is desirable to have an EL system learn to improve its annotation performance over time, our approach can be used to generate training data with an acceptable quality for weakly supervised methods. Moreover, accurate entity links form the basis for more elaborate IE tasks. E.g. for relation extraction they answer the question *between which entities does this relation hold?* and for sentiment analysis the question *who expresses this sentiment about what?*

## 6. CONCLUSIONS

The current state-of-the-art entity linking systems aim to be open-domain solutions for corpora that are as heterogeneous as the Web. An unfortunate effect of this aim is that such generalist EL systems often disappoint when they are used on domain-specific corpora. We have proposed and evaluated a solution that is highly cost-effective in comparison with existing alternative approaches.

We have outlined the prerequisites for, and development of, a lightweight linking system that targets salient entity types in a specific corpus. In our approach, the output of such specialist linkers is combined in a simple manner with that of an off-the-shelf EL system, which is responsible for linking mentioned entities of non-salient types that are also of interest to the corpus' users. The specialist system, two baseline generalist systems, and hybrid combinations thereof have been evaluated against a gold standard that has been carefully constructed by two human annotators who have experience in using the selected corpus. This gold standard, along with system annotations, annotation guidelines and accompanying code, is available as an open-data benchmark for the EL community at *http://datahub.io/dataset/el-bm-nl-9912*.

Our results show that the specialist system offers competitive performance to the two baseline systems, even though it is limited to two highly specific entity types. Moreover, by combining the specialist linkers with one or both generalist EL systems, recall can be significantly increased at a modest precision cost.

**Acknowledgements** This research was supported by the Netherlands Organization for Scientific Research (ExPoSe project, NWO CI # 314.99.108; DiLiPaD project, NWO Digging into Data # 600.006.014). We extend our gratitude to Evelijn Martinius and Rosa Merino Claros for helping to prepare the gold standard.

## 7. REFERENCES

[1] Aho, A. V. and Corasick, M.J. 1975. Efficient string matching: an aid to bibliographic search. *Communications of the ACM*. 18, (1975), 333–340.

[2] Berlanga, R., Nebot, V. and Pérez, M. 2014. Tailored semantic annotation for semantic search. *Journal of Web Semantics*. (2014), 1–13.

[3] Van den Bosch, A., Busser, B., Canisius, S. and Daelemans, W. 2007. An efficient memory-based morphosyntactic tagger and parser for Dutch. *Selected Papers of CLIN 2007* (Leuven, Belgium, 2007), 99–114.

[4] Daiber, J., Jakob, M., Hokamp, C. and Mendes, P.N. 2013. Improving Efficiency and Accuracy in Multilingual Entity Extraction. *Proc. of I-Semantics 2013* (Austria, Graz, 2013), 3–6.

[5] Daumé III, H., Kumar, A. and Saha, A. 2010. Frustratingly Easy Semi-Supervised Domain Adaptation. *Proceedings of DANLP '10* (2010), 53–59.

[6] Gagnon, M., Zouaq, A. and Jean-Louis, L. 2013. Can we use linked data semantic annotators for the extraction of domain-relevant expressions? *Proc. of WWW 2013 companion* (2013), 1239–1246.

[7] De Goede, B., Marx, M., Nusselder, A. and van Wees, J. 2011. Succinct summaries of narrative events using social networks. *Proc. of HT '11*. (2011), 299–304.

[8] Odijk, D., Meij, E. and de Rijke, M. 2013. Feeding the Second Screen: Semantic Linking based on Subtitles. *OAIR 2013* (2013).

[9] Piskorski, J. and Yangarber, R. 2013. Information Extraction: Past, Present, and Future. *Multi-Source, Multilingual Information Extraction and Summarization*. Springer-Verlag. 23–50.

[10] Shen, W., Wang, J. and Han, J. 2014. Entity Linking with a Knowledge Base: Issues, Techniques, and Solutions. *IEEE Transactions on Knowledge and Data Engineering*. 4347, 2 (2014), 443–460.

[11] Usbeck, R. et al. 2015. GERBIL – General Entity Annotator Benchmarking Framework. *Proc. of WWW 2015* (Florence, Italy, 2015).